\documentclass[10pt]{iopart}

\usepackage{graphicx}
\usepackage{upgreek}
\usepackage{xcolor}

\begin{document}
\title[Integrated SNSPDs on titanium in-diffused lithium niobate waveguides]
{Integrated superconducting nanowire single-photon detectors on titanium in-diffused lithium niobate waveguides}

\author{Jan Philipp H{\"o}pker$^{*1}$, Varun B. Verma$^2$, Maximilian Protte$^{1}$, Raimund Ricken$^1$, Viktor Quiring$^1$, Christof Eigner$^3$, \textcolor{black}{Lena Ebers$^4$, Manfred Hammer$^4$, Jens F{\"o}rstner$^4$,} Christine Silberhorn$^3$, Richard P. Mirin$^2$, Sae Woo Nam$^2$, and Tim J. Bartley$^1$}
\address{$^1$ Mesoscopic Quantum Optics, Department of Physics, Paderborn University, Warburger Str. 100, 33098 Paderborn, Germany \\
$^2$ National Institute of Standards and Technology, 325 Broadway, Boulder, Colorado 80305, USA \\
$^3$ Integrated Quantum Optics, Department of Physics, Paderborn University, Warburger Str. 100, 33098 Paderborn, Germany \\
\textcolor{black}{$^4$ Theoretical Electrical Engineering, Department of Electrical Engineering and Information Technology, Paderborn University, Warburger Str. 100, 33098 Paderborn, Germany}}
\ead{$^*$jan.philipp.hoepker@upb.de}
\vspace{10pt}
\begin{indented}
\item[]April 2021
\end{indented}

\begin{abstract}
We demonstrate the integration of amorphous tungsten silicide superconducting nanowire single-photon detectors on titanium in-diffused lithium niobate waveguides. We show proof-of-principle detection of evanescently-coupled photons of 1550\,nm wavelength using bidirectional waveguide coupling for two orthogonal polarization directions. We investigate the internal detection efficiency as well as detector absorption using coupling-independent characterization measurements. Furthermore, we describe strategies to improve the yield and efficiency of these devices.
\end{abstract}
%
%
%
%
\ioptwocol

\section{Introduction}
The fundamental constituents of all quantum photonics experiments are the ability to generate, manipulate and measure single photons very precisely. These components are typically combined in quantum optical circuits, the size of which scale with the complexity of task to be performed~\cite{Wang2020}. A crucial strategy for scaling these devices is therefore developing photonic integration platforms which are capable of performing these tasks. Many integration platforms exist, and they can be broadly categorized as based on semiconductor (including but not limited to Si, SiN, AlN, InP), crystalline (e.g. LiNbO$_{3}$, KTP) or glass (SiO$_{2}$) substrates~\cite{Bogdanov2017}. Each platform offers different advantages and disadvantages with respect to intrinsic guiding efficiency, fabrication tolerance, component density and potential nonlinear properties enabling active components~\cite{Elshaari2020}.

Of these platforms, titanium in-diffused waveguides in congruently-grown lithium niobate is a highly promising platform for quantum communication applications~\cite{Sharapova2017}. At telecom wavelengths, these waveguides support two orthogonal modes with losses below 0.03\,dB/cm, as well as direct end-face coupling to SMF-28 fiber with interface losses $<$1\,dB~\cite{Montaut2017}. The high second-order nonlinear properties of lithium niobate, combined with quasi-phasematching by periodic poling, enable a range of nonlinear interactions relevant to quantum photonics~\cite{Luo2019}, including photon-pair generation~\cite{Montaut2017}, frequency conversion, and spectral-temporal mode manipulation~\cite{Ansari2018}. Furthermore, its high electro-optic coefficient enables direct manipulation of phase, spatial modes (through tunable directional couplers) and polarization at high speed with very low energy dissipation~\cite{Thiele2020}. 

An additional important component required for a fully-integrated quantum circuit on lithium niobate are efficient evanescently-coupled single-photon detectors. Such detectors have been deposited on lithium niobate~\cite{Tanner2012,Hoepker2017,Smirnov2018,Agruzov2019}, as was as on waveguides based on lithium niobate on insulator (LNOI)~\cite{Colangelo2020, Sayem2020, Lomonte2021}. However, evanescent coupling to detectors deposited onto titanium in-diffused waveguides in lithium niobate has not previously been demonstrated. For these waveguides, high-efficiency detection can still be achieved off-chip with fiber coupling. However, there are some applications, for example feed-forward~\cite{Prevedel2007}, \textcolor{black}{where it is necessary to minimize the physical distance between detectors and active optoelectronic circuitry to minimize latency.}

Superconducting nanowire single-photon detectors (SNSPDs) have become a standard tool in quantum photonics experiments, due to their high efficiency~\cite{Reddy2020}, low noise~\cite{Hochberg2019} and excellent timing resolution~\cite{Korzh2020}. These detectors have enabled a range of groundbreaking experiments in quantum optics (and beyond)~\cite{Shalm2015}, many of which have also made use of photonic integrated circuits for generation and/or manipulation of photons~\cite{Wang2020}. Moreover, these detectors have been directly deposited on a range of quantum photonic integration platforms~\cite{Ferrari2018}. 

Several different superconducting materials have been shown to be responsive to single photons~\cite{Natarajan2012,Holzman2019,You2020}. Of these, amorphous superconductors such as WSi and MoSi are expected to be ideally suited to integration on lithium niobate, since there are no lattice constants to be matched to the underlying substrate.

In this paper, we report on results and challenges when integrating amorphous WSi SNSPDs on titanium in-diffused waveguides in congruently grown lithium niobate. We show that the detectors respond as expected when deposited on the substrate, and show a broad plateau, indicating saturation of the internal detection efficiency. We verify coupling to the evanescent field of the waveguide by exploiting bidirectional coupling to the detector from each polarization mode. Nevertheless, the intrinsic pyroelectric properties of lithium niobate currently limit the yield of these devices, and the large optical mode size relative to the small detector area currently limits the on-chip efficiency of these devices.

The manuscript is organized as follows. We first present the fabrication of the waveguides and subsequent detector deposition. We present results of the detector response, the system detection efficiency and system jitter. We also discuss the yield and challenges with dissipation of pyroelectric charges, and finish with an outlook outlining possible mitigation strategies. 

\section{Fabrication}
\begin{figure}[htbp]
  \centering
  \includegraphics[width=0.49\textwidth]{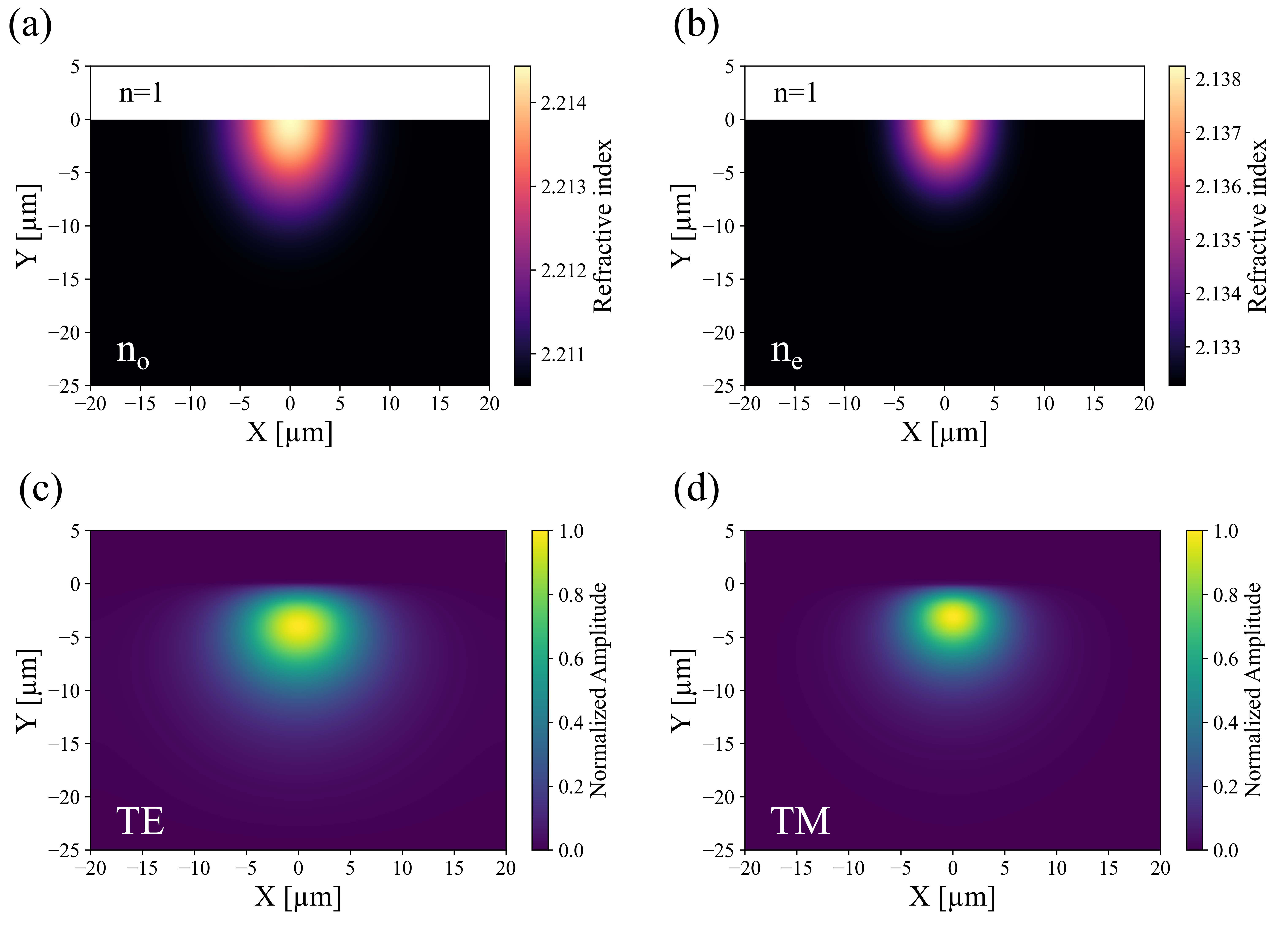}
\caption{Modelled refractive index distribution of titanium in-diffused waveguides at 1550\,nm wavelength for light in TE polarization \textcolor{black}{seeing the ordinary refractive index} in (a) and TM polarization \textcolor{black}{seeing the extraordinary refractive index} in (b). Corresponding simulated waveguide modes are shown in (c) and (d).}
\label{Fig1}
\end{figure}

Waveguide fabrication begins by depositing 80\,nm titanium on a congruent lithium niobate wafer with e-beam evaporation, followed by a positive photoresist. Titanium stripes of width 5\,$\upmu$m, 6\,$\upmu$m, and 7\,$\upmu$m which will form the waveguides are patterned with vacuum contact lithography and subsequent wet etching. The remaining titanium stripes diffuse into the lithium niobate substrate at 1060\,$^{\circ}$C which creates a slowly-varying refractive index profile from 2.211 to 2.214 in transverse electric (TE) polarization and 2.133 to 2.138 in transverse magnetic (TM) polarization at 1550\,nm wavelength. These profiles based on Ref.~\cite{Bartnick2021}, along with the resulting modes, are shown in Figure~\ref{Fig1}.

Following the waveguide patterning, the wafer section is diced and its end-faces are polished to produce a chip of length 2.3\,cm with 70 waveguides. Prior to further processing, linear losses are characterized using the interferometric technique described by Regener and Sohler~\cite{Regener1985}. These waveguides support both TE and TM modes with losses below 0.03\,dB/cm.

\begin{figure}[htbp]
  \centering
  \includegraphics[width=0.49\textwidth]{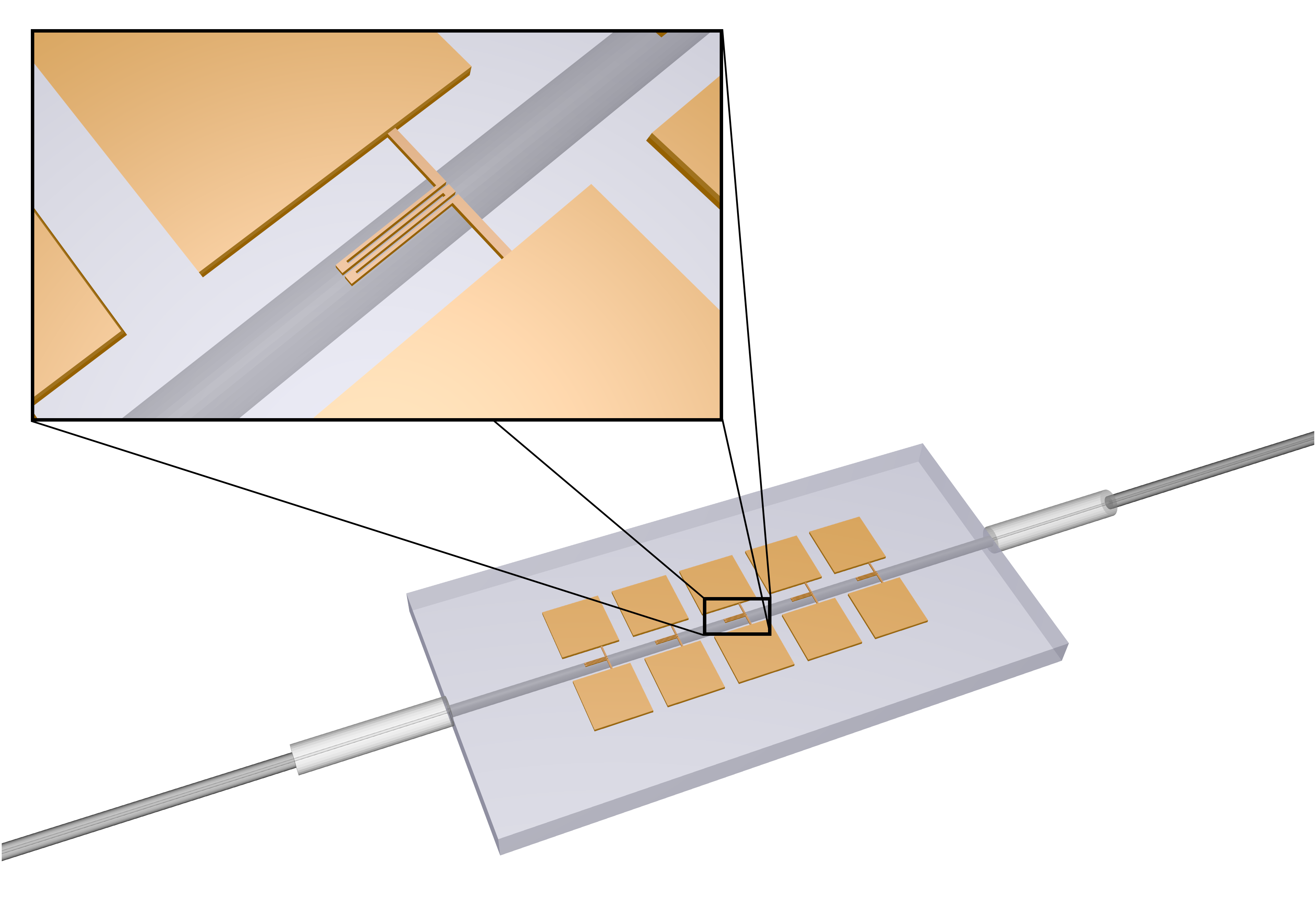}
\caption{Schematic of the fiber-coupled waveguide-chip with five in-line SNSPDs. The detector's ``w''-shape geometry with 400\,$\upmu$m length and 160\,nm wire width is visualized in the inset.}
\label{Fig2}
\end{figure}

Following the loss measurements, deposition and subsequent patterning of the superconducting film is conducted. This comprises a homogeneous 5\,nm thick layer of amorphous tungsten silicide \textcolor{black}{and a 2\,nm thick silicon capping layer}, deposited using magnetron sputtering. The layer is then structured using electron beam lithography to define the nanowire meander structures, followed by optical lithography for the contact electrodes. This results in a ``w''-shape geometry of 160\,nm width, 400\,$\upmu$m length, and 160\,nm spacing, as shown in Figure~\ref{Fig2}. \textcolor{black}{The cross-section of the device is 4$\times$5\,nm$\times$160\,nm. This is significantly smaller than the mode size shown in Figure~\ref{Fig1}. Nevertheless, absorption can occur along the entire length of the device. A total of five detectors on one waveguide were deposited.}

\subsection{Loss measurements after detector deposition}
After detector deposition, the waveguide losses are measured again using the same reversible characterization scheme by Regener and Sohler~\cite{Regener1985}. Waveguides with five in-line SNSPDs show a transmission loss increase to 0.10\,dB/cm $\pm$0.03\,dB/cm in TE and 0.05\,dB/cm $\pm$0.03\,dB/cm in TM. Attributing the loss increase completely to photon-absorption in the detectors would suggest an absorption efficiency per detector of 0.6\% in TE and 0.2\% in TM. \textcolor{black}{These numbers agree with two-dimensional finite element method simulations, where detection efficiencies of 1\% for the TE mode and 0.4\% for the TM mode were calculated for our geometry. Besides a small refractive index change at cryogenic temperatures which can be included into the loss characterization scheme, no temperature dependent loss was observed within our measurement accuracy.}

\subsection{Optical connection}
The Ti in-diffused waveguides are weakly guiding, and as such support spatial modes with a mode field diameter which closely matches that of single-mode fiber. This enables us to ``pigtail'' the device by butt-coupling a fiber ferrule directly to the end-face of the chip. To do so, the fiber ferrule is optically aligned to couple light through the waveguide using high-precision motorized stages. Once positioned, a small dose of index-matched, UV-cured optical adhesive is applied between the ferrule and chip face. This is precured using a ring of UV LEDs placed around the joint, followed by a UV gun to finish hardening the glue.

\textcolor{black}{The maximum mode overlap of the fiber and waveguide modes is 92\%, which places an upper bound on the achievable coupling efficiency. At room temperature, the maximum achieved coupling efficiency in practice at this interface is 84\%~\cite{Montaut2017}.} This will be further reduced if no antireflective coatings are used at the interface. \textcolor{black}{In the absence of dielectric coatings,} we regularly achieve more than 70\% coupling using this approach.

For use in the cryostat, it is crucial to both minimize the amount of glue used and aim for a uniform thickness at the interface~\cite{Hoepker2019}. We note that the glue is not specified to function below $-150$\,$^{\circ}$C. \textcolor{black}{For these reasons,} optical coupling efficiency is variable: \textcolor{black}{the maximum cryogenic throughput through both fiber connections we have achieved is as high as 66\%~\cite{Hoepker2017b}, however the throughput can change from one temperature cycle to the next, due to a variety of irreversible cryogenic effects (for example pyro- and piezoelectric discharge, outgassing from the adhesive, plastic-phase material contraction etc). Nevertheless,} the coupling efficiency per interface estimated for this chip was only 26\% $\pm$3\% while it was 48\% $\pm$1\% at room temperature. Further refinements to this procedure are necessary to improve the consistency of cryogenic operation.

\section{Optical response}
\begin{figure}[htbp]
  \centering
  \includegraphics[width=0.49\textwidth]{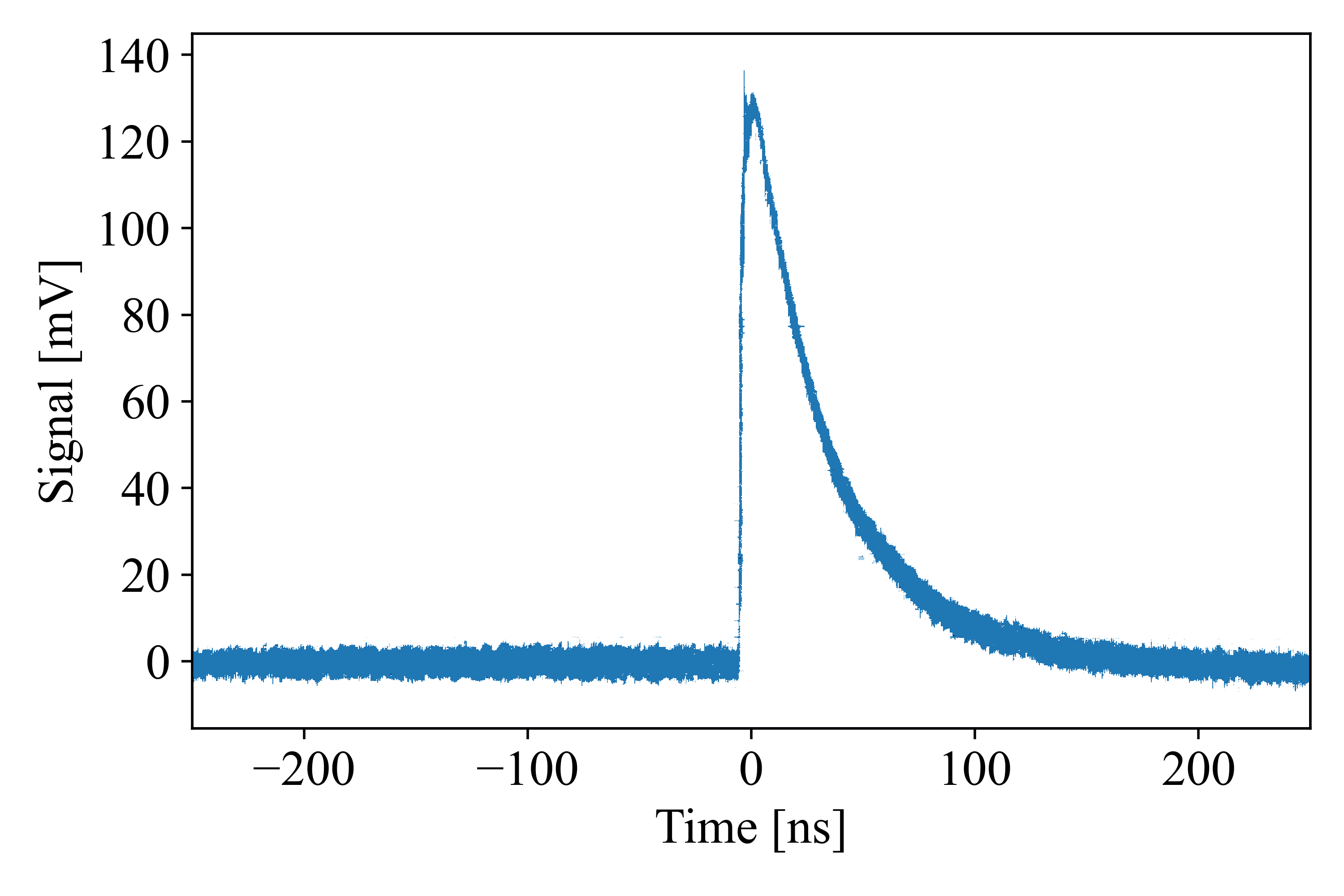}
\caption{Measured detector response of an integrated SNSPD.}
\label{Fig3}
\end{figure}

Prior to testing evanescent-field coupling, the photosensitivity of the device was tested under flood illumination \textcolor{black}{in a closed-cycle sorption cryostat at 0.8\,K,} using a pulsed laser of 1556.3\,nm wavelength, 9\,ps pulse width and 16\,ps timing jitter to the synchronization output. Of the five detectors deposited, two showed a response as shown in Figure~\ref{Fig3}. \textcolor{black}{The other three were found to be open after cooling down.} 

\subsection{Internal efficiency}
\begin{figure}[htbp]
  \centering
  \includegraphics[width=0.49\textwidth]{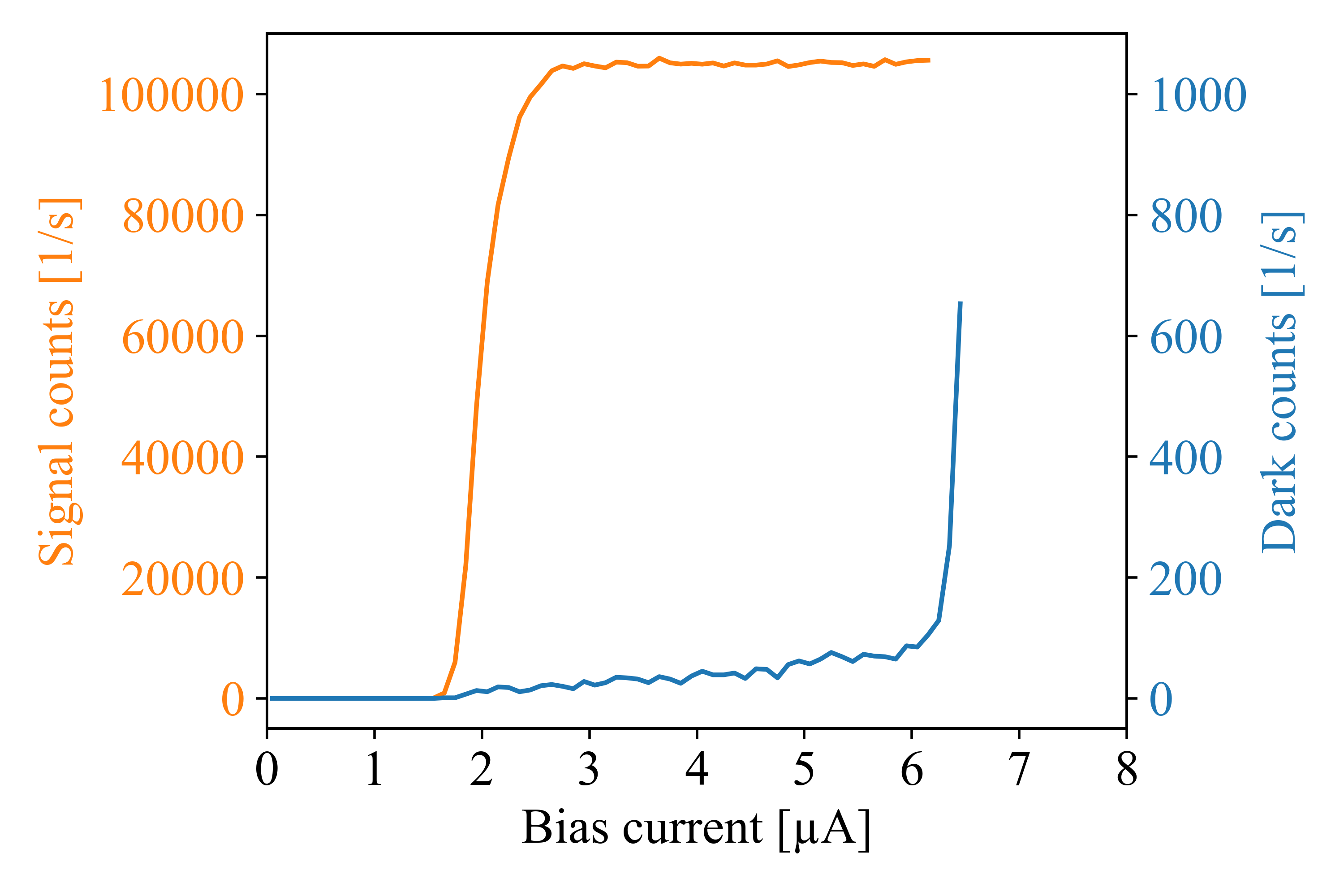}
\caption{Measured signal counts of an integrated SNSPD showing a saturation plateau, as well as noise counts (dark counts) for different detector bias currents.}
\label{Fig4}
\end{figure}

This configuration also allows us to verify saturating the internal detection efficiency and dark counts of the device. To do so, the bias current is varied between 0\,$\upmu$A and 7\,$\upmu$A while counts are monitored. Saturation of these counts as the bias current increases indicates that the internal detection efficiency is maximized, as can clearly be seen in Figure~\ref{Fig4}. 

\subsection{System detection efficiency}
Following the flood illumination tests, we can test the optical response to light coupled through the waveguide. To do so, four separated measurements of the system detection efficiency are conducted, corresponding to two independent coupling directions, and two independent polarizations. This also allows us to estimate the on-chip efficiency for each device to be 0.7\% in one polarization and 0.1\% in the other, since the bidirectional fiber coupling allows for a coupling-independent efficiency measurement.

\textcolor{black}{We measured a system jitter of 380\,ps, providing an upper limit on the detector jitter itself. This is higher in comparison to other detectors of similar geometry~\cite{Marsili2013}, possibly due to additional electronics such as long wire bonds which limit high-speed operation. Furthermore, the laser and time-tagger jitter are not deconvoluted.}

\textcolor{black}{Although scattered photons can still contribute to the on-chip efficiency, evanescent coupling could still be verified. Measurement events due to evanescent coupling scale linearly with coupling efficiency, whereas the probability of measuring scattered photons decreases with coupling efficiency. Furthermore, the rate of measuring scattered photons decreases sharply with the distance from the scattering center, here the fiber connection. However, evanescently-coupled photons are much less sensitive to position. We investigated these effects using multiple in-line detectors and bidirectional coupling, as well as exploiting the polarization-dependent coupling and detector absorption, to distinguish measurements due to scatter and evanescent coupling. Each detector along the waveguide showed similar count rates, independent of its position. This suggests that scattered photons from the input fiber to waveguide interface do not contribute significantly to the measured count rates.}

\section{Optimizing efficiency and yield}
\begin{figure}[htbp]
  \centering
  \includegraphics[width=0.49\textwidth]{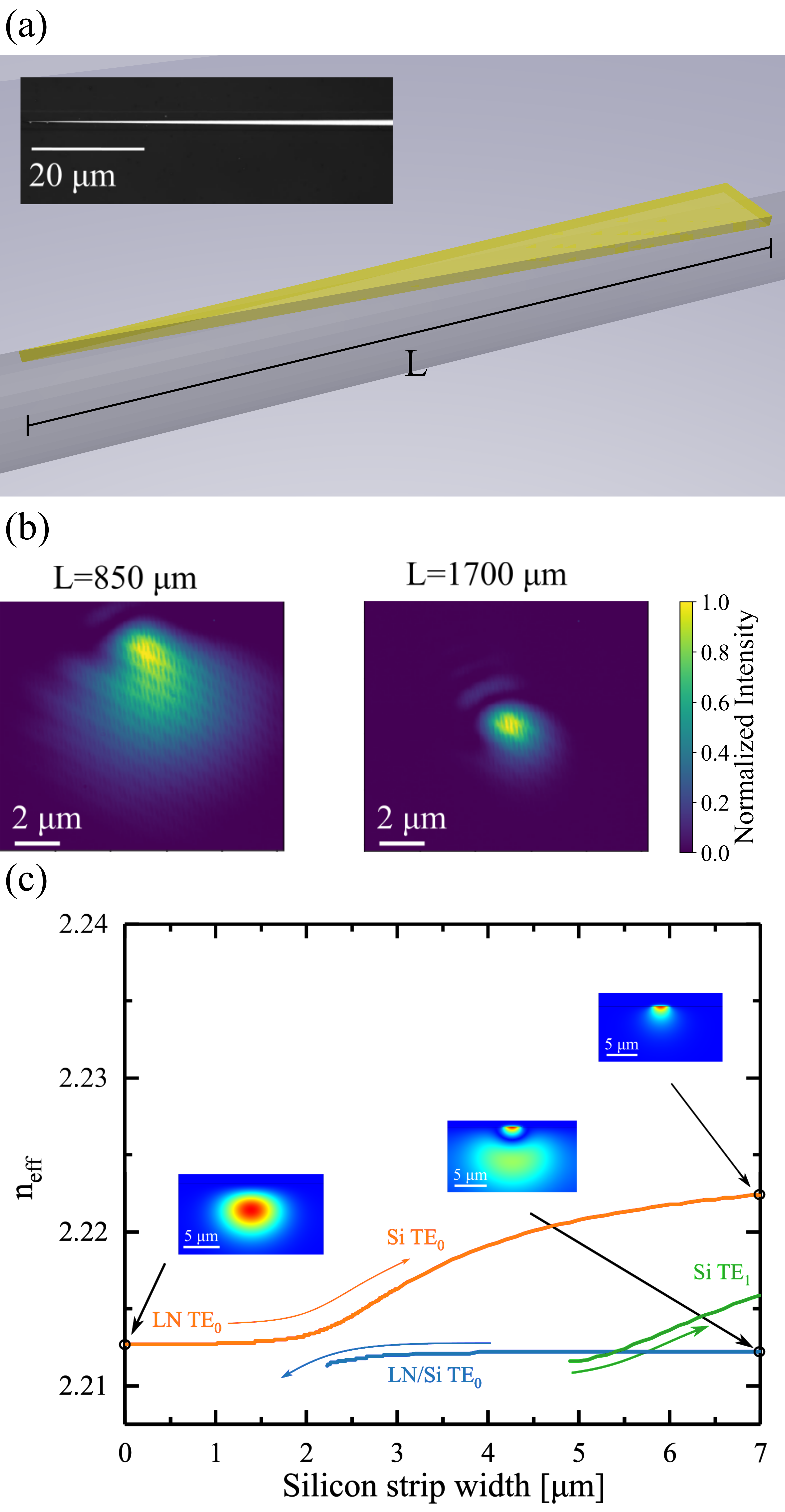}
\caption{(a) Schematic of a silicon taper structure atop the dark-grey waveguide which can be used for enhancing the light confinement in the detector region. The inset shows a micrograph of a fabricated structure. (b) Measured mode images at the end-position of the taper for two different taper length. \textcolor{black}{(c) Simulation results for the effective refractive index calculated for different modes over the width of the silicon taper.}}
\label{Fig5}
\end{figure}

While these results show integrating SNSPDs on Ti in-diffused lithium niobate is possible in practice, it is clear there are many challenges before optimized devices can be fabricated. One major challenge is the on-chip coupling efficiency, which is significantly limited by the overlap of the optical mode with the detector. This may be overcome by using taper structures to manipulate the optical mode in the region of the detectors~\cite{Parfenov2020}. We have taken initial steps in this direction, with the fabrication of Si-tapers atop the titanium in-diffused waveguides, as shown in Figure~\ref{Fig5}.

For this, a 60\,nm thick silicon film is deposited using PECVD and structured using laser lithography. Stripes with a tapered width from 0\,$\upmu$m to 3\,$\upmu$m and varying length up to 1700\,$\upmu$m are fabricated and covered with a protective silicon dioxide layer. Our \textcolor{black}{theoretical simulations} and measurements show that this can \textcolor{black}{significantly} decrease the mode size, as seen in Figure~\ref{Fig5}, depending on the \textcolor{black}{taper} length to at least 36\% of its initial size, corresponding to the resolution limit of our measurement setup~\cite{Protte2020}. Further simulations of these structures have shown an estimated coupling efficiency to the detector of above 90\% using optimized tapers. The next step is to combine these taper structures with the integrated detectors to verify the increase in efficiency.

A second approach \textcolor{black}{which we will pursue in future} is simply to extend the length of the device along the waveguide. While E-beam written nanowires are limited by relatively small write fields, recent results have shown that superconducting detectors of this kind may be patterned using optical lithography~\cite{Chiles2020, Charaev2020}. As such, detectors of lengths up to several mm become possible.

\subsection{Pyroelectric damage}
A second major challenge is to improve the yield for these devices during the cooldown. We believe this is limited by the pyroelectric properties of lithium niobate. This effect causes charges to accumulate on the surface of the chip~\cite{Jachalke2017}; \textcolor{black}{the particular surface where charges accumulate} depends on the crystal cut with respect to the crystal axis. We use z-cut lithium niobate since this will allow fabrication of periodic domain inversion, necessary for quasi-phasematching various nonlinear optical processes. Unfortunately, this cut means that the charges accumulate on the top surface of the chip, where the waveguides and detectors are located.

\begin{figure}[htbp]
  \centering
  \includegraphics[width=0.49\textwidth]{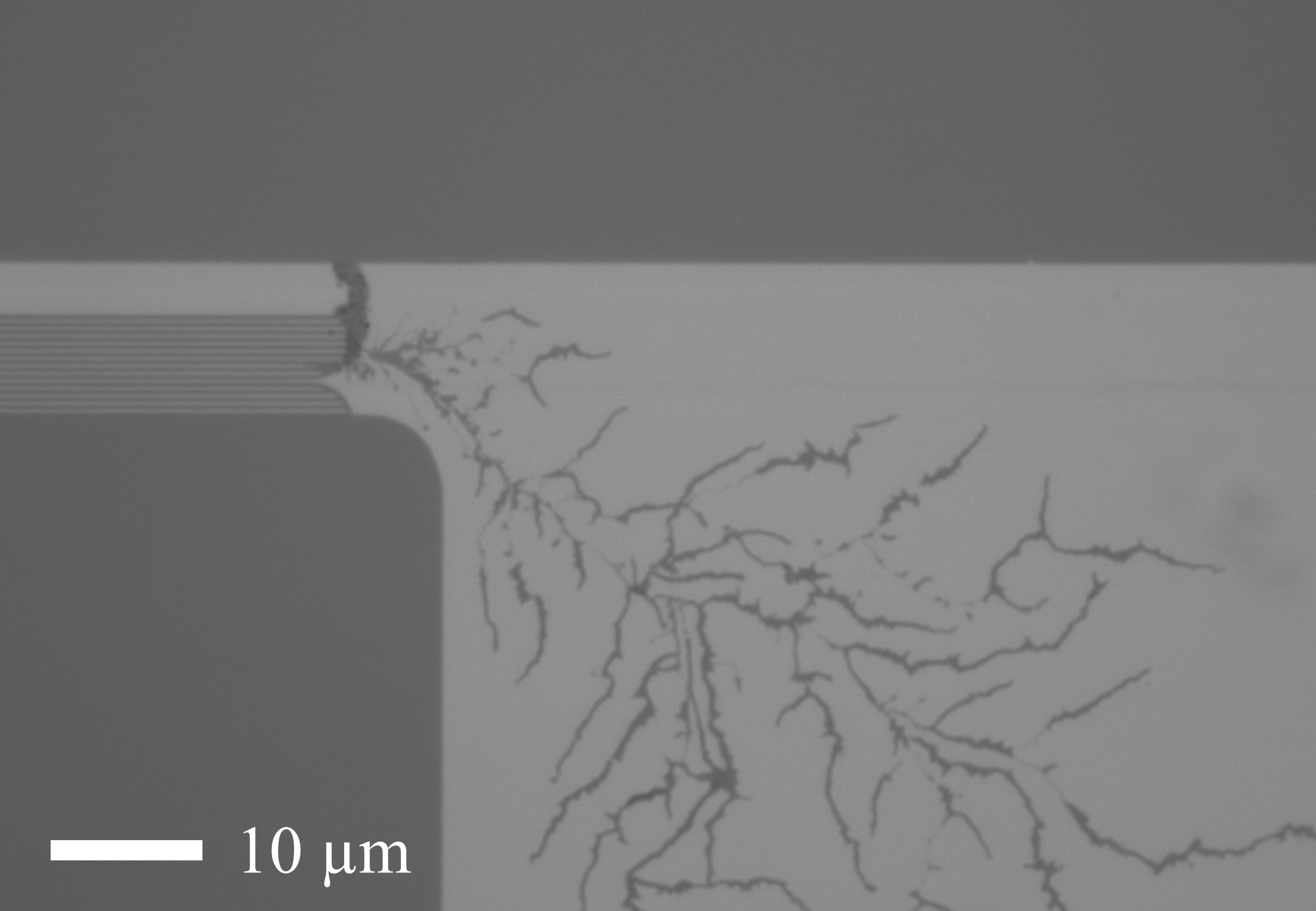}
\caption{Micrograph of a damaged SNSPD after one temperature cycle.}
\label{Fig6}
\end{figure}

Under ambient conditions, these charges dissipate by recombining with atmospheric ions. Under vacuum, there is no obvious pathway for these charges to dissipate. Placing conductive materials on top of the device may aid the dissipation, however one must be careful to maintain the low-loss optical guiding properties. In the case of the SNSPDs it appears that in some cases, these fields caused by the pyroelectric charges can cause discharge across the highly resistive nanowire region, before it reaches temperatures at which it becomes superconductive. These discharges can cause significant damage to the film, as shown in Figure~\ref{Fig6}.

The effect does not always happen, therefore further optimization is required to find structures which do not suffer from this effect. Indeed, this effect was not observed when integrating a different flavor of superconducting detector on lithium niobate, namely the transition edge sensor~\cite{Hoepker2019}.

This suggests that the physical size of the conductor plays a role, which may mean that the larger-area structures may prove beneficial here too.

\section{Conclusion}
In summary, integrating superconducting nanowire single-photon detectors with titanium in-diffused lithium niobate wavegudies is necessary for a number of quantum photonic circuits on this platform. We have presented the first proof-of-principle results that evanescent coupling to these detectors is possible, however a number of challenges remain before this becomes more widely adopted. We described some of the challenges and mitigation strategies associated with optimizing yield and on-chip coupling efficiency. Furthermore, this system offers several advantages such as high fiber-coupling efficiency, independent detector absorption verification, and cryogenic compatibility with other second-order nonlinear optical processes required for quantum photonic processing.

\section*{Acknowledgements}
The study was partially funded by the Deutsche Forschungsgemeinschaft (DFG, German Research Foundation) - Projektnummer No. 231447078—TRR 142 and Bundesministerium f{\"u}r Bildung und Forschung (13N14911); European Research Council
(725366).

\section*{References}
\bibliographystyle{iopart-num}
\bibliography{references}

\end{document}